\documentclass[aip,apl,amsmath,amssymb,preprint,reprint]{revtex4-1}

\usepackage[dvipsnames]{xcolor}
\usepackage{siunitx}
\usepackage{graphicx} 
\usepackage{dcolumn}
\usepackage{bm}
\usepackage[utf8]{inputenc}
\usepackage[T1]{fontenc}
\usepackage{mathptmx}
\usepackage{etoolbox}

\makeatletter
\def\@email#1#2{%
 \endgroup
 \patchcmd{\titleblock@produce}
  {\frontmatter@RRAPformat}
  {\frontmatter@RRAPformat{\produce@RRAP{*#1\href{mailto:#2}{#2}}}\frontmatter@RRAPformat}
  {}{}
}%
\makeatother

\begin{document}

\preprint{AIP/123-QED}

\title{Feedback-enabled magnomechanics in lithium ferrite}

\author{Mehri Ebrahimi}
\affiliation{Department of Physics, University of Alberta,
Edmonton, Alberta T6G 2E9, Canada}

\author{Yunhu Huang}
\affiliation{Department of Physics, University of Alberta,
Edmonton, Alberta T6G 2E9, Canada}

\author{Ali Rashedi}
\affiliation{Department of Physics, University of Alberta,
Edmonton, Alberta T6G 2E9, Canada}

\author{John P. Davis}
\email{jdavis@ualberta.ca}
\affiliation{Department of Physics, University of Alberta,
Edmonton, Alberta T6G 2E9, Canada}

\date{\today}

\begin{abstract}
Lithium ferrite (LiFe) is a promising material for cavity magnonics because its large spin density enables strong coupling between magnons and microwave photons. Its potential for cavity magnomechanics, however, has remained unexplored. Here, we observe magnomechanical interactions in a single-crystal LiFe sphere using coherent microwave feedback to suppress dissipation of the cavity--magnon polariton. In the absence of sufficient feedback, the narrow mechanical response is difficult to resolve against the much broader polariton background. Increasing the feedback gain reduces the polariton linewidth and correspondingly increases the magnomechanical cooperativity, revealing a clear magnomechanically induced transparency feature. In the measurements presented here, the effective upper-polariton linewidth is reduced from $3.53~\mathrm{MHz}$ without feedback to $5.8~\mathrm{kHz}$ in the presence of feedback, while the measured cooperativity increases from $C=1.9\times10^{-3}$ to $C=0.15$. These measurements provide, to our knowledge, the first observation of cavity magnomechanics in LiFe and demonstrate coherent feedback as a practical route for accessing weak interactions that would otherwise be obscured by dissipation.

\end{abstract}

\maketitle

\section{Introduction}

Hybrid magnonic systems combine collective spin excitations with electromagnetic and mechanical degrees of freedom and provide a versatile platform for studying coherent interactions between otherwise distinct excitations. In cavity magnonics, the magnetic-dipole interaction between magnons and microwave photons can readily enter the strong-coupling regime, producing hybrid cavity--magnon polaritons~\cite{Soykal2010, Huebl2013, Tabuchi2014, Zhang2014, Goryachev2014, Bourhill2016, Flower2019, Rameshti2022, Ebrahimi2026CPA, Chumak2022}. These systems have enabled studies of coherent information transfer, non-Hermitian dynamics, microwave signal processing, and sensing \cite{Xu2021, Harder2021, Ebrahimi2021, Engelhardt2022, Yao2023, Gui2026, Kounalakis2023, Xu2024}.

Most cavity-magnonics experiments have employed yttrium iron garnet (YIG), in large part because of its low magnetic dissipation. Other ferrimagnetic materials can nevertheless offer complementary properties. Lithium ferrite (LiFe, $\mathrm{LiFe_5O_8}$) has long been developed as a microwave ferrite, with a high Curie temperature, high electrical resistivity, and comparatively narrow ferromagnetic-resonance linewidths \cite{WhitePatton1978,Pachauri2015}. More recently, LiFe has emerged as an attractive cavity-magnonics material because its small unit cell gives a high spin density and therefore a large collective magnon--photon coupling. Goryachev \textit{et al.} demonstrated strong coupling between sub-millimeter LiFe spheres and three-dimensional microwave cavities at millikelvin temperatures, observing coupling rates up to $250~\mathrm{MHz}$ at $9.5~\mathrm{GHz}$ with magnon linewidths of order $4~\mathrm{MHz}$ \cite{GoryachevLiFe2018}. These properties establish LiFe as a promising alternative material platform for cavity magnonics, while leaving its cavity-magnomechanical behavior largely unexplored.

Mechanical motion provides an additional degree of freedom through magnetoelastic interactions. In cavity magnomechanics, magnetostriction couples the magnon population to mechanical deformation of the magnetic resonator, analogous to radiation-pressure coupling in cavity optomechanics. Under a strong microwave drive, this interaction can be parametrically enhanced and detected through phenomena such as magnomechanically induced transparency (MMIT), the magnetic analogue of optomechanically induced transparency \cite{Weis2010,ZhangMagnomechanics2016,Potts2021}. Experiments to date have used YIG, where low magnetic dissipation helps make the weak magnomechanical interaction experimentally accessible \cite{ZhangMagnomechanics2016,Potts2021}.

Extending cavity magnomechanics to other magnetic materials is challenging because the observable response depends not only on the intrinsic magnetoelastic interaction but also on dissipation of the hybrid mode participating in the interaction. In particular, a broad cavity--magnon polariton can obscure the much narrower mechanical response even when the system is driven strongly. This suggests a route to accessing otherwise hidden magnomechanical interactions, through engineering the effective dissipation of the polariton. Similar feedback concepts have been developed extensively in cavity optomechanics, where feedback can reshape susceptibilities, enhance cooling, and even enable normal-mode splitting \cite{Rossi2017,Rossi2018,Zippilli2018,Ernzer2023}. This technique is well suited to cavity magnomechanics because the magnomechanical cooperativity scales inversely with the polariton linewidth, allowing exploration of magnomechanics in otherwise impossible materials. Recently, we demonstrated using YIG that coherent microwave feedback can strongly modify the dissipation of a cavity--magnon polariton and thereby access interaction regimes that are otherwise inaccessible \cite{EbrahimiFeedback2026}. 

In this work, we use coherent microwave feedback to observe cavity magnomechanics in a single-crystal LiFe sphere. Feedback gain progressively narrows the cavity--magnon polariton and reveals a narrow MMIT feature near the mechanical resonance. We reach a magnomechanical cooperativity of $C=0.15$, modest compared to YIG but nonetheless promising. These measurements establish LiFe as a cavity-magnomechanical material and show that active dissipation engineering can expose weak interactions that are otherwise hidden by dissipation.


\section{Results and Discussion}

\begin{figure}[t]
    \centering
    \includegraphics[width=0.85\columnwidth]{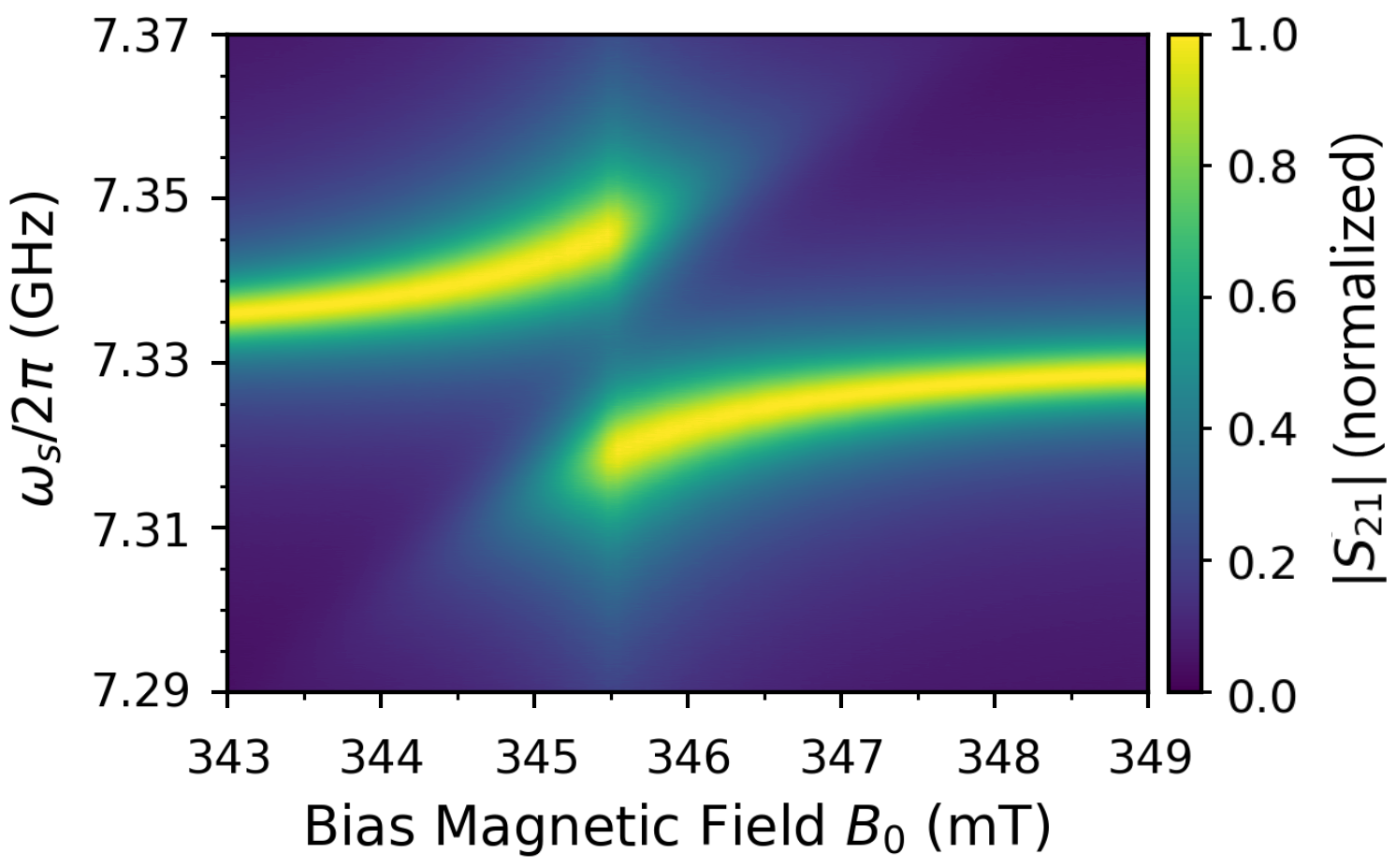}
    \caption{Normalized transmission spectrum $|S_{21}|$ as a function of bias magnetic field $B_0$, showing the avoided crossing between the cavity and magnon modes. The resulting upper and lower branches correspond to the hybridized cavity--magnon polariton modes.}
    \label{fig1}
\end{figure}

The experiment uses a $200~\mu\mathrm{m}$-diameter single-crystal LiFe sphere positioned in a three-dimensional copper microwave cavity. The cavity supports a $\mathrm{TE}_{101}$ mode at $\omega_a/2\pi=7.33~\mathrm{GHz}$, while the frequency of the uniform magnon mode is tuned with an external magnetic field, $B_0$, according to $\omega_m=\gamma B_0$, where $\gamma/2\pi=28~\mathrm{GHz/T}$ is the gyromagnetic ratio. The bare cavity and magnon linewidths are $\kappa_a/2\pi=4.08~\mathrm{MHz}$ and $\kappa_m/2\pi=8.77~\mathrm{MHz}$, respectively.  The magnetic-dipole interaction hybridizes the cavity and magnon modes to form upper and lower cavity--magnon polaritons, as shown by the avoided crossing shown in Fig.~\ref{fig1}, with a magnon--photon coupling strength of $g_{ma}/2\pi=13.10~\mathrm{MHz}$.

\begin{figure}[t]
    \centering
    \includegraphics[width=0.86\columnwidth]{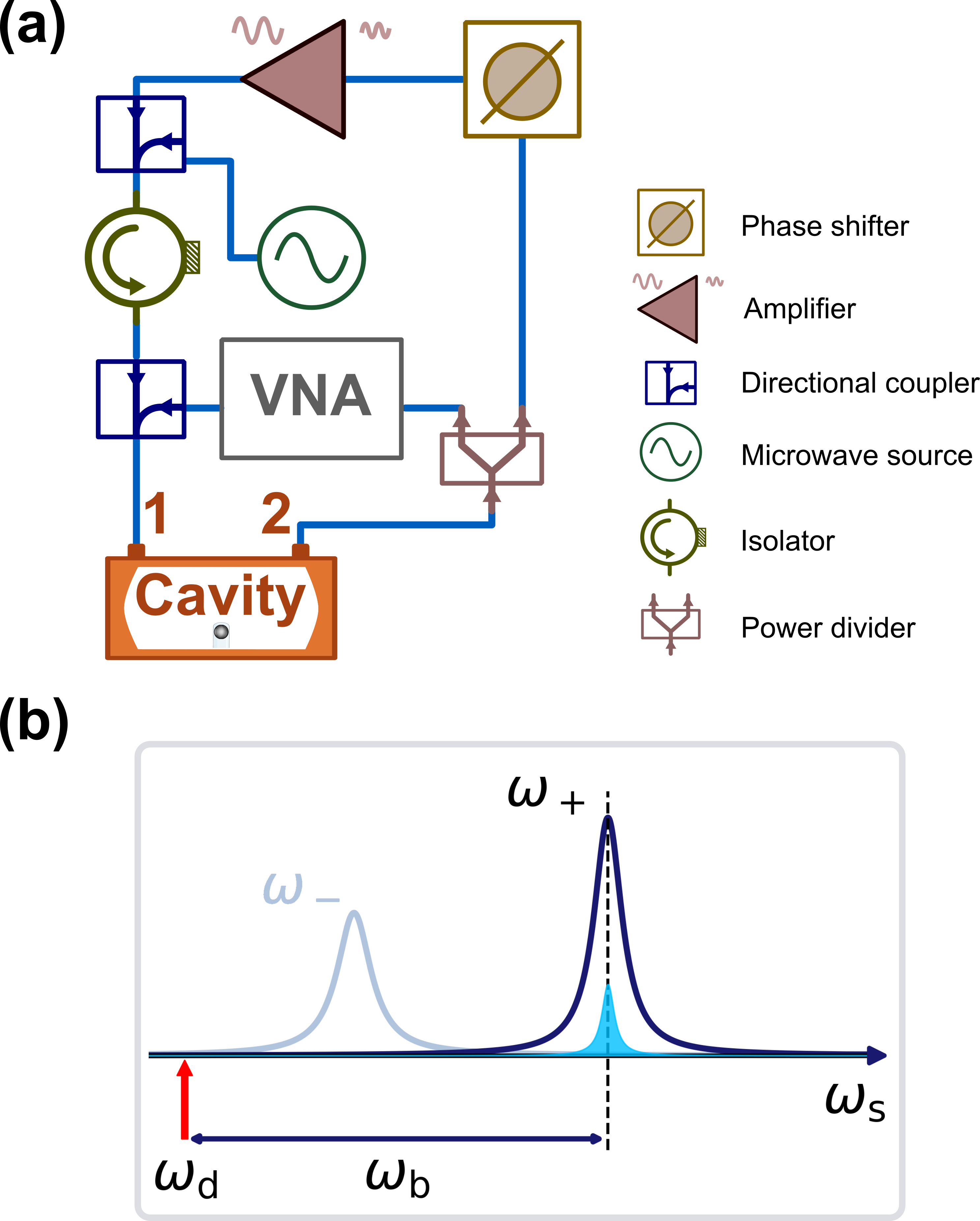}
    \caption{
(a) Schematic of the cavity magnomechanical system and microwave feedback loop. The LiFe sphere is positioned inside a copper microwave cavity and probed by a vector network analyzer (VNA). A fraction of the cavity output is routed through the feedback loop, where the signal is amplified, phase shifted, and subsequently fed back into the cavity. An independent microwave source provides the drive tone for the magnomechanical measurements. 
(b) Frequency schematic of the measurement showing the drive tone, $\omega_d$, approximately one mechanical frequency, $\omega_b$, red-detuned from the upper-polariton mode, $\omega_+$.
}

    \label{fig2}
\end{figure}

Coherent microwave feedback is implemented using the same general approach as in our previous work.\cite{EbrahimiFeedback2026} A fraction of the microwave field leaving the cavity is amplified, phase shifted, and coherently returned to the cavity input (Fig.~\ref{fig2}(a)). The feedback modifies both the resonance frequency and dissipation rate of the microwave mode. The frequency and dissipation are therefore modified to be, 
\begin{eqnarray}
\tilde{\omega}_a
&=&
\omega_a+
2\sqrt{\kappa_{\mathrm{ext}}^{(1)}\kappa_{\mathrm{ext}}^{(2)}}
 g_{\mathrm{fb}}\sin\phi,
\\
\tilde{\kappa}_a
&=&
\kappa_a-
2\sqrt{\kappa_{\mathrm{ext}}^{(1)}\kappa_{\mathrm{ext}}^{(2)}}
 g_{\mathrm{fb}}\cos\phi,
\end{eqnarray}
where $g_{\mathrm{fb}}$ and $\phi$ are the feedback gain and phase, and $\kappa_{\mathrm{ext}}^{(1)}$ and $\kappa_{\mathrm{ext}}^{(2)}$ are the external decay rates through ports 1 and 2 of the cavity, respectively.~\cite{EbrahimiFeedback2026} The feedback phase is chosen to maximize linewidth suppression, while the magnetic field is adjusted as necessary to maintain the desired cavity--magnon operating point.

Because the cavity and magnon modes are strongly coupled, the feedback-induced modification of the cavity parameters directly modifies the properties of the hybrid cavity--magnon polariton modes. In particular, when the magnon mode is tuned into resonance with the feedback-modified cavity mode, $\omega_m=\tilde{\omega}_a$, the effective polariton linewidths are given by
\begin{equation}
\tilde{\kappa}_{\pm}
=
\frac{\tilde{\kappa}_a+\kappa_m}{2},
\end{equation}
where $\kappa_m$ is the intrinsic magnon dissipation rate. Thus, by tuning the feedback gain and phase to reduce $\tilde{\kappa}_a$, the dissipation of the hybrid polariton modes can be directly suppressed. The optimum feedback phase is $\phi=0$, which, for a fixed feedback gain, provides the maximum reduction of the polariton linewidths. In this regime, the gain introduced by the feedback loop partially compensates the intrinsic dissipation of the cavity--magnon system, resulting in substantially narrower polariton resonances.

We probe the magnomechanical interaction using a strong red-detuned microwave drive and a weak probe tone. The drive is applied near the red mechanical sideband of the upper polariton, $\omega_d=\omega_+-\omega_b$, where the relevant LiFe mechanical resonance lies near $\omega_b/2\pi=15.23~\mathrm{MHz}$ (Fig.~\ref{fig2}(b)). In this configuration, the parametrically enhanced interaction produces a small MMIT feature within the much broader upper-polariton response.

\begin{figure}[t]
    \centering
    \includegraphics[width=1\columnwidth]{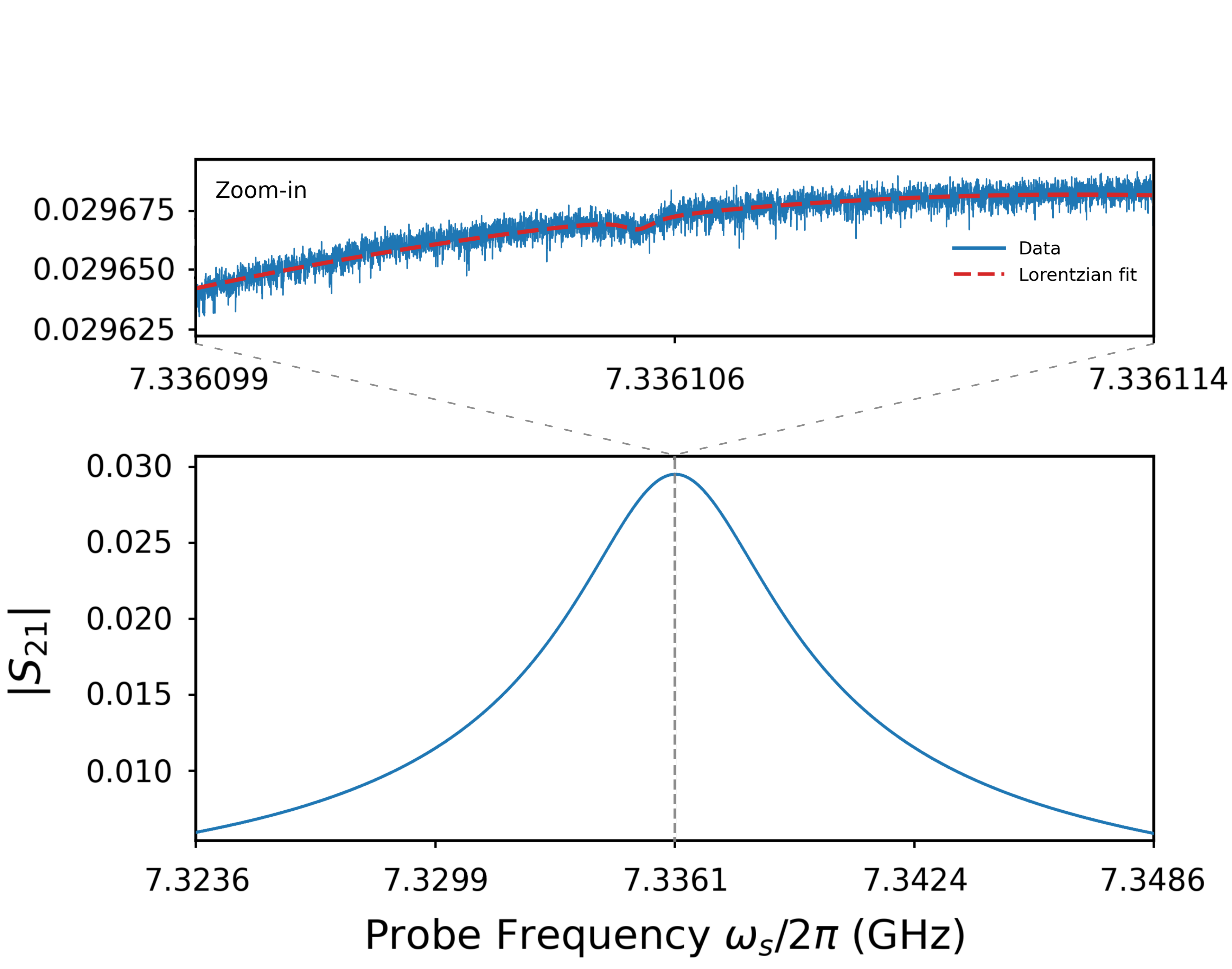}
    \caption{ Transmission amplitude $|S_{21}|$ of the upper polariton measured without feedback at a drive power of $25~\mathrm{dBm}$.  The expanded view in the upper reveals the weak magnomechanical feature near the mechanical resonance, which is barely detectable against the broader upper-polariton response.}

    \label{fig3}
\end{figure}

We first characterize the magnomechanical response in the absence of feedback. At a drive power of $25~\mathrm{dBm}$, the upper-polariton response is shown in Fig.~\ref{fig3}. The magnomechanical feature is extremely weak and barely detectable within the polariton resonance. The expanded view reveals only a small feature near the mechanical resonance, demonstrating the weak magnomechanical cooperativity in LiFe without feedback, even under strong microwave driving.

\begin{figure}[b]
    \centering
    \includegraphics[width=0.48\textwidth]{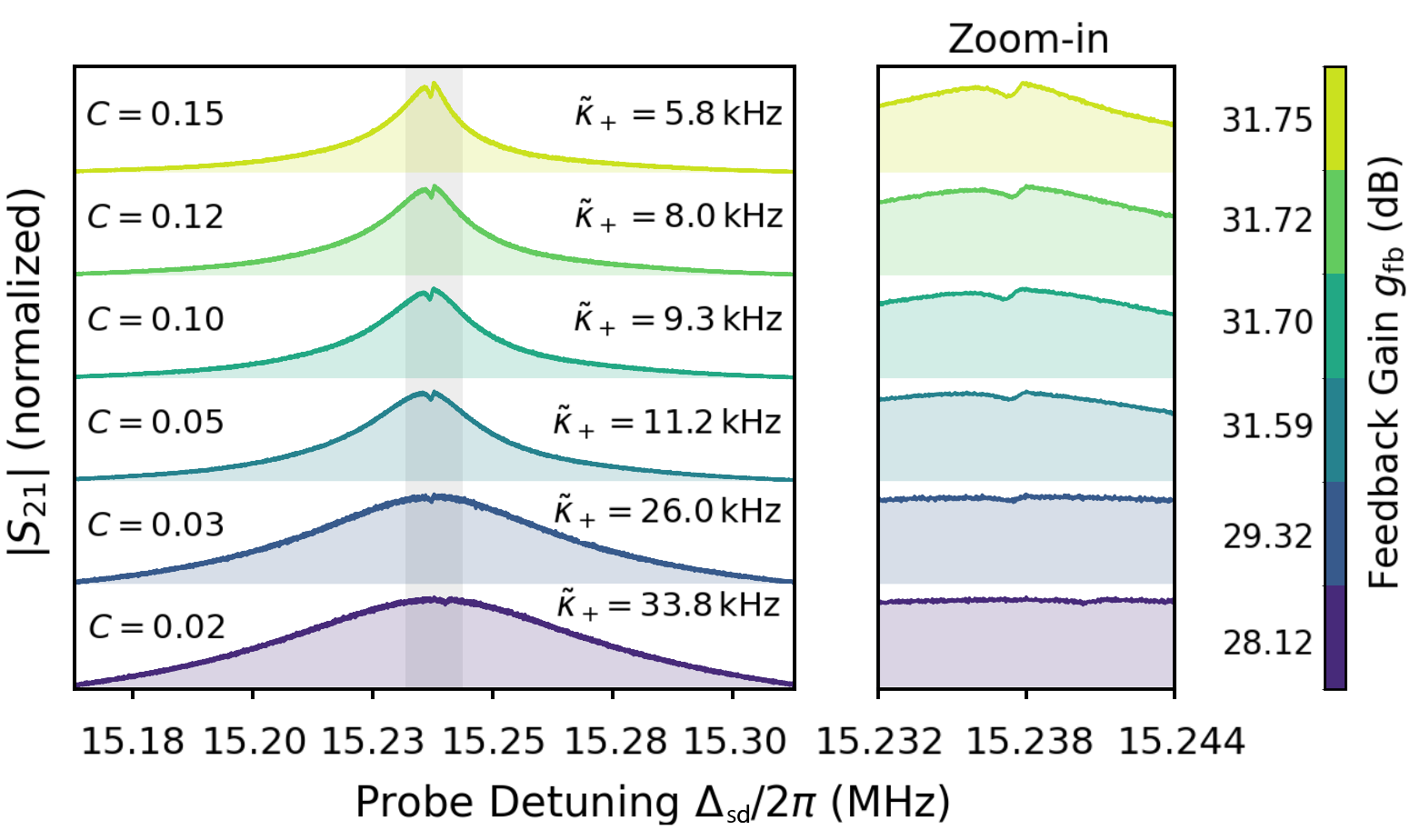}
    \caption{Feedback-enabled observation of magnomechanically induced transparency in LiFe. Normalized transmission $|S_{21}|$ is shown as a function of probe detuning $\Delta_{sd}=\omega_s-\omega_d$ for increasing feedback gain $g_{\mathrm{fb}}$ (bottom to top), with constant pump tone power, $P_d = 22.6\,\mathrm{dBm}$. The feedback suppresses the upper-polariton linewidth from $\kappa_+/2\pi = 3.53~\mathrm{MHz}$ without feedback to $\tilde{\kappa}_+/2\pi = 33.8~\mathrm{kHz}$ and $5.8~\mathrm{kHz}$ at feedback gains $g_{\mathrm{fb}} = 28.12~\mathrm{dB}$ and $31.75~\mathrm{dB}$, respectively, while increasing the extracted magnomechanical cooperativity from $C = 1.9\times10^{-3}$ without feedback to $C = 0.02$ and $0.15$, respectively. Curves are vertically offset for clarity and are color coded by feedback gain. The right panel enlarges the region around the mechanical resonance, showing the narrow MMIT feature that becomes increasingly visible as the polariton linewidth is reduced.}
    \label{fig4}
\end{figure}

\begin{figure}[t]
    \centering
    \includegraphics[width=0.90\columnwidth]{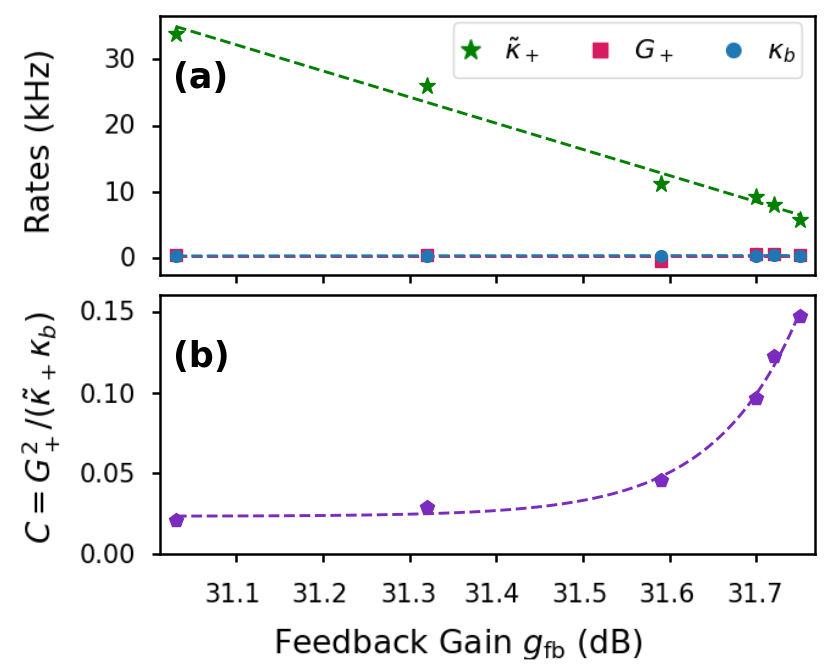}
    \caption{Effect of coherent feedback on the magnomechanical interaction. (a) Extracted effective upper-polariton linewidth $\tilde{\kappa}_+$, drive-enhanced upper-polariton--phonon coupling rate $G_+$, and mechanical linewidth $\kappa_b$ as functions of the feedback gain $g_{\mathrm{fb}}$. Increasing the feedback gain strongly reduces $\tilde{\kappa}_+$, while $G_+$ and $\kappa_b$ remain comparatively small. Dashed lines are guides to the eye. (b) Corresponding magnomechanical cooperativity $C=G_+^2/(\tilde{\kappa}_+\kappa_b)$ as a function of feedback gain. The suppression of the upper-polariton linewidth results in a pronounced increase in the cooperativity. The dashed line is a guide to the eye.}
    \label{fig5}
\end{figure}

Figure \ref{fig4} shows instead that the MMIT feature becomes resolvable, and the cooperativity enhanced, through  feedback gain. The normalized transmission $|S_{21}|$ is plotted as a function of probe detuning for a sequence of feedback gains. At the lowest feedback gain, the upper-polariton linewidth is $\tilde{\kappa}_+/2\pi=33.8~\mathrm{kHz}$ and the extracted cooperativity is only $C=0.02$ (see Fig.~\ref{fig5}). The narrow mechanical feature is correspondingly difficult to distinguish from the polariton background. As the feedback gain is increased, the polariton narrows and the mechanical feature becomes progressively more visible. At the highest feedback gain $\tilde{\kappa}_+/2\pi=5.8~\mathrm{kHz}$ and the cooperativity reaches $C=0.15$ (Fig.~\ref{fig5}).

The effect of feedback can be understood directly through the magnomechanical cooperativity, 
\begin{equation}
C=\frac{G_+^2}{\kappa_b\tilde{\kappa}_+},
\label{eq:cooperativity}
\end{equation}
where $G_+$ is the drive-enhanced upper-polariton--phonon coupling rate, $\kappa_b$ the mechanical linewidth, and $\tilde{\kappa}_+$ effective upper-polariton linewidth. For fixed mechanical dissipation and comparable drive conditions, reducing $\tilde{\kappa}_+$ increases the cooperativity and therefore the contrast of the MMIT response. The data in Fig.~\ref{fig5} show this behavior directly: the approximately sixfold reduction in the polariton linewidth is accompanied by a substantial increase in $C$ and by the transition from a barely discernible mechanical signature to a clearly resolved transparency feature.

The feedback-enabled spectra can also be used to extract the underlying single-magnon magnomechanical coupling rate $g_{mb}$. Following the theory of our previous work, ~\cite{EbrahimiFeedback2026} we extract $g_{mb}/2\pi=0.26~\mathrm{mHz}$ for the LiFe sphere. This provides a quantitative characterization of cavity magnomechanics in LiFe and enables a direct comparison with YIG, for which single-magnon magnomechanical coupling rates in millimeter- and sub-millimeter-scale spheres have been characterized experimentally ~\cite{ZhangMagnomechanics2016,Potts2021,Bittencourt2023}. For a YIG sphere of the same size, we obtain $g_{mb}/2\pi=4.83~\mathrm{mHz}$ following the same theoretical approach, showing that the single-magnon magnomechanical coupling in LiFe is substantially weaker than that in YIG, which was not known a priori. The relevant magnomechanical parameters for LiFe and YIG are summarized in Table~I. In our previous YIG experiment with feedback~\cite{EbrahimiFeedback2026}, the larger single-magnon magnomechanical coupling rate, together with more favorable dissipation parameters, allowed coherent feedback to drive the system into the strong polariton--mechanical coupling regime. In LiFe, by contrast, the substantially smaller $g_{mb}$ and less favorable dissipation parameters make this regime considerably more difficult to access. Nevertheless, the strong feedback-induced suppression of the polariton linewidth is sufficient to enhance the cooperativity and render the otherwise weak magnomechanical interaction experimentally detectable.

These measurements suggest a direction for future experiments. The magnetic properties of LiFe are temperature dependent, including its magnetocrystalline anisotropy \cite{Stelmashenko1966}, while temperature-dependent magnetic damping is well established in YIG \cite{MaierFlaig2017,Huang}. Recent measurements in YIG/GGG acoustic resonators have additionally reported a temperature dependence of the magnon--phonon coupling \cite{Weber2026}. These observations motivate a direct study of the temperature dependence of the single-magnon magnomechanical coupling in LiFe. Because feedback can compensate, at least in part, for changes in polariton dissipation, it offers a way to separate a variation in $g_{mb}(T)$ from the changing visibility caused by magnetic dissipation. A comparative temperature-dependent study of LiFe and YIG is therefore a reasonable extension of the present work.

\begin{table}[t]
\caption{Comparison of the single-magnon magnomechanical coupling rate and characteristic dissipation rates for LiFe and YIG spheres. Here, $g_{mb}$ is the single-magnon magnomechanical coupling rate, $\kappa_{+}$ and $\tilde{\kappa}_{+}$ denote the upper-polariton linewidths without and with feedback, respectively, and $\kappa_b$ is the mechanical linewidth.}
\label{tab:LiFe_YIG_comparison}
\centering

\renewcommand{\arraystretch}{1.15}

\begin{tabular*}{\columnwidth}{@{\extracolsep{\fill}} c c c}
\hline\hline
Parameter & LiFe & YIG \\
\hline
$g_{mb}/2\pi$ (mHz)               & 0.26 & 4.83 \\
$\kappa_{+}/2\pi$ (MHz)          & 3.53 & 2.71 \\
$\tilde{\kappa}_{+}/2\pi$ (kHz)  & 5.80 & 0.30 \\
$\kappa_{b}/2\pi$ (kHz)          & 0.50 & 2.06 \\
\hline\hline
\end{tabular*}

\end{table}


\section{Conclusion}

We have observed cavity magnomechanics in a single-crystal LiFe sphere using coherent microwave feedback. Without feedback gain the mechanical response is difficult to resolve because the narrow magnomechanical feature is obscured by the broader cavity--magnon polariton. Increasing the feedback gain suppresses the polariton linewidth from $3.53~\mathrm{MHz}$ without feedback to $5.8~\mathrm{kHz}$ in the presence of feedback and increases the magnomechanical cooperativity to $C=0.15$, producing a clearly resolved MMIT feature. To our knowledge, these measurements constitute the first observation of cavity magnomechanics in LiFe.

More broadly, these results demonstrate coherent feedback as a practical tool for accessing weak interactions in dissipative hybrid systems. Rather than relying exclusively on materials with intrinsically narrow resonances, feedback permits the effective susceptibility and dissipation of a hybrid mode to be engineered externally. This approach may enable comparative measurements of magnomechanical coupling across a broader range of magnetic materials and their temperature dependence.

\begin{acknowledgments}
The authors acknowledge that the land on which this work was performed is on Treaty Six Territory, the traditional lands of many First Nations, Métis, and Inuit in Alberta. They acknowledge the support from the Natural Sciences and Engineering Research Council, Canada (Grant Nos.~RGPIN-2022-03078,  ALLRP 592535-23, ALLRP 568609-21); Alberta Innovates (Grant No.~212200780); the National Research Council (Grant QSP-017-1); and the University of Alberta Department of Physics through the Advah Bhatia Fellowship.
\end{acknowledgments}

\bibliography{references_life}

\end{document}